\documentclass[prb,twocolumn,showpacs,superscriptaddress]{revtex4}

\usepackage[dvips]{graphicx}
\usepackage{amsmath}

\begin{document}
\title{Berezinskii-Kosterlitz-Thouless Type Scenario in Molecular Spin Liquid $A$Cr$_2$O$_4$}

\author{M.~Hemmida}
\affiliation{Experimental Physics V, Center for Electronic
Correlations and Magnetism, University of Augsburg, 86135 Augsburg,
Germany}

\author{H.-A.~Krug~von~Nidda}
\affiliation{Experimental Physics V, Center for Electronic
Correlations and Magnetism, University of Augsburg, 86135 Augsburg,
Germany}

\author{V.~Tsurkan}
\affiliation{Experimental Physics V, Center for Electronic Correlations and Magnetism, University of Augsburg, 86135 Augsburg, Germany}
\affiliation{Institute of Applied Physics, Academy of Sciences of Moldova, MD-2028 Chisinau, Republic of Moldova}

\author{A.~Loidl}
\affiliation{Experimental Physics V, Center for Electronic
Correlations and Magnetism, University of Augsburg, 86135 Augsburg,
Germany}

\date{\today}

\begin{abstract}

The spin relaxation in chromium spinel oxides $A$Cr$_{2}$O$_{4}$ ($A=$ Mg, Zn, Cd) is investigated in the paramagnetic regime by electron spin resonance (ESR). The temperature dependence of the ESR linewidth indicates an unconventional spin-relaxation behavior, similar to spin-spin relaxation in the two-dimensional (2D) chromium-oxide triangular lattice antiferromagnets. The data can be described in terms of a generalized Berezinskii-Kosterlitz-Thouless (BKT) type scenario for 2D systems with additional internal symmetries. Based on the characteristic exponents obtained from the evaluation of the ESR linewidth, short-range order with a hidden internal symmetry is suggested.

\end{abstract}



\maketitle


\section{Introduction}

The search of a molecular spin-liquid state in strongly frustrated spinels attracts enormous theoretical and experimental interest in the condensed-matter community.\cite{Tchernyshyov2002,Lee2000,Lee2002} A spin molecule is a spin cluster that is spatially confined within a geometrical shape such as an atomic molecule in which the intramolecular correlation dominates the intermolecular one.\cite{Tomiyasua2008} In the normal spinel structure $AB_{2}$O$_{4}$ the pyrochlore lattice of the $B$-site occupied by magnetic ions with antiferromagnetic interaction provides the geometrically frustrated framework for the formation of spin molecules. Experimental evidence has been indicated, e.g. based on neutron-scattering studies.

Focusing on inelastic neutron-scattering (INS) results in spinels, the form factor in ZnCr$_{2}$O$_{4}$ was convincingly interpreted in terms of antiferromagnetic multi-spin clusters or precisely hexagonal loops (hexamers), made up of six tetrahedra where two spins of each tetrahedron occupy the vertices of the hexamers. The remaining two spins of a given tetrahedron belong to a different hexamer.\cite{Lee2000,Lee2002} INS measurements on single crystals of MgCr$_{2}$O$_{4}$ did not only show the presence of hexamers in the antiferromagnetic phase, but also additional seven-spin clusters (heptamers), formed by the spins of two corner-sharing tetrahedra.\cite{Tomiyasua2008} Very recent INS measurements in the same compound revealed the existence of hexa- and heptamers also in the paramagnetic phase.\cite{Tomiyasu2013} The spin correlations in the paramagnetic phase are interpreted as a paramagnetic scattering of lowest geometric quantum modes or geometrons. Moreover, powder INS studies in HgCr$_{2}$O$_{4}$ exhibit molecular-type excitations in form of dodecamers consisting of two hexamers.\cite{Tomiyasu2011} Similar multi-spin clusters (oligomers) have been observed in many other spinel compounds like dimers and octamers in CuIr$_{2}$S$_{4}$,\cite{Radaelli2002} heptamers in AlV$_{2}$O$_{4}$,\cite{Matsuda2006} hexamers in NiCr$_{2}$O$_{4}$ and FeCr$_{2}$O$_{4}$,\cite{Tomiyasub2008a} tetramers and di-tetramers in GeCo$_{2}$O$_{4}$,\cite{Tomiyasu2011c,Watanabe2011} hexamers and dodecamers in ZnFe$_{2}$O$_{4}$.\cite{Tomiyasub2011a}

From theoretical point of view, Tchernyshyov \textit{et al.} considered a model in which spin rearrangements around such hexagons are dominant.\cite{Tchernyshyov2002} They discuss a mechanism for lifting the frustration through a coupling between spin and lattice degrees of freedom. The high symmetry of the pyrochlore lattice and the spin degeneracy drive a distortion of tetrahedra via a magnetic Jahn-Teller ("spin-Teller") effect. The resulting state exhibits a reduction from cubic to tetragonal symmetry and the development of bond order in the spin system with unequal two-spin correlations $\langle\textbf{S}_{\rm i}\cdot\textbf{S}_{\rm j}\rangle$ on different bonds of a tetrahedron. The spectrum of spin excitations in the distorted antiferromagnet contains a large number of modes clustered near a finite frequency. These magnons, a remnant of pyrochlore zero modes, are confined to strings of parallel spins, so called string modes. The string modes can live on straight lines of parallel spins, spirals, irregular lines, or even closed loops. A mode living on a short loop is a local resonance.

The formation of such spin loops strongly reminds to the configuration of the magnetic vortices in two-dimensional (2D) magnets. This was originally described by Berezenskii,\cite{Berezinskii1972} Kosterlitz and Thouless \cite{Kosterlitz1973} for the classical vortices with $U(1)$-symmetry in the XY model and later by Kawamura and Miyashita \cite{Kawamura1984} for vortices with $Z_{2}$-symmetry in the triangular Heisenberg antiferromagnet. In both types of 2D magnets the spin-spin relaxation turned out to be governed by the presence of the vortices resulting in a characteristic temperature dependence of the electron spin resonance (ESR) linewidth $\Delta H$. In the paramagnetic regime the influence of vortices on the linewidth was approximately derived as $\Delta H \propto \xi^{3}$ where $\xi$ denotes the vortex correlation length.\cite{Becker1996} Experimental realization was found, e.g., in the 2D XY antiferromagnet BaNi$_2$V$_2$O$_8$ (Ref.~\onlinecite{Heinrich2003}) and in triangular Heisenberg antiferromagnets $A$CrO$_2$ with $A=$ H, Li, K, Cu, Ag, and Pd (Refs.~\onlinecite{Hemmida2009, Hemmida2011}). Here we will show that in $A$Cr$_{2}$O$_{4}$ the ESR line broadening can be described in terms of a generalized vortex scenario, suggesting that the linewidth provides a direct access to the string modes in the paramagnetic phase.

\section{Experimental Details}

Polycrystalline ZnCr$_{2}$O$_{4}$, MgCr$_{2}$O$_{4}$, and CdCr$_{2}$O$_{4}$ samples were prepared by solid-state reaction from high-purity binary oxides in evacuated quartz ampoules. The synthesis was
repeated several times in order to reach good homogeneity. Magnetization measurements have been performed using a superconducting quantum interference device (SQUID) MPMS5 (Quantum Design) at temperatures $2 \leq T \leq 300$~K. The ESR measurements were performed in a Bruker ELEXSYS E500-CW spectrometer at X-band (9.4 GHz) frequency equipped with a continuous He-gas flow cryostat (Oxford Instruments) working in the temperature range $4.2 \leq T \leq 300$~K. The polycrystalline samples were fixed in a quartz tube with paraffin. Due to the lock-in technique with field modulation the field derivative of the microwave--absorption signal is detected as a function of the static magnetic field. Resonance absorption occurs, if the incident microwave energy matches the energy of magnetic dipolar transitions between the electronic Zeeman levels.

\section{Results}

\begin{figure}
\centering
\includegraphics[width=75mm]{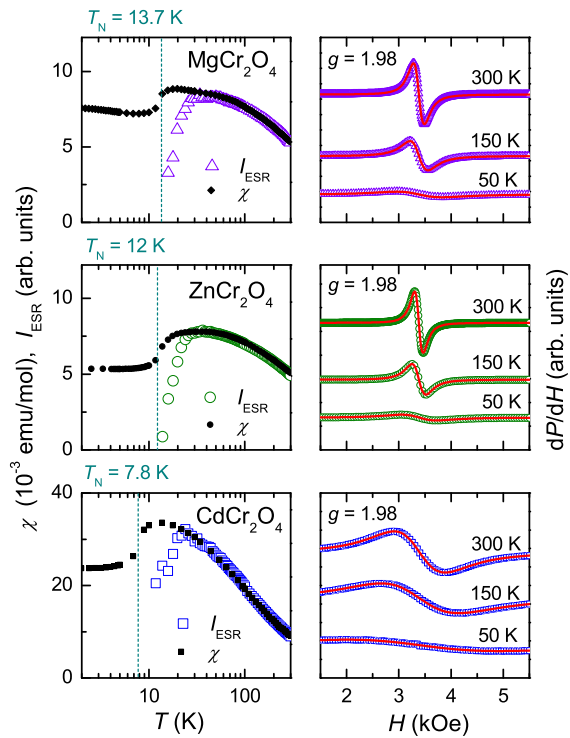}
\caption{(Color online) Left: temperature dependence of the ESR intensities and SQUID susceptibilities of MgCr$_{2}$O$_{4}$, ZnCr$_{2}$O$_{4}$, and CdCr$_{2}$O$_{4}$. Right: ESR spectra in X-band for selected temperatures in the paramagnetic regime. The solid line indicates the fit with the field derivative of a Lorentz line.}
\label{Spectra}
\end{figure}

\begin{figure}
\centering
\includegraphics[width=75mm]{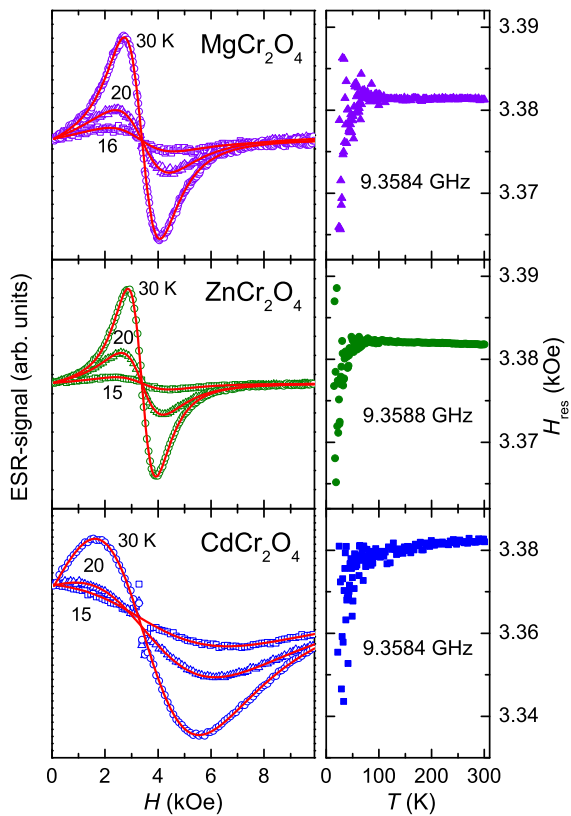}
\caption{(Color online) Left: selected ESR spectra close to $T_{\rm N}$ (solid lines indicate fits with the field derivative of a Lorentz line) and Right: temperature dependence of paramagnetic resonance fields of MgCr$_{2}$O$_{4}$, ZnCr$_{2}$O$_{4}$, and CdCr$_{2}$O$_{4}$. at X-band frequency}
\label{SpectraLT}
\end{figure}

In the whole paramagnetic regime, above the N\'eel temperature $T_{\rm N}$, the ESR spectra of all three compounds, shown in the right frames of Fig.~\ref{Spectra}, are well described by the field derivative of single Lorentz lines following
\begin{equation}
\frac{dP}{dH} = A \cdot \frac{-2x}{(1+x^2)^2}, \hspace{0.25cm} {\rm with} \hspace{0.25cm} x = \frac{H-H_{\rm res}}{\Delta H}  \label{Lorentz}
\end{equation}
Here $H_{\rm res}$ denotes the resonance field, $\Delta H$ the half width at half maximum (HWHM), and $A$ the amplitude factor. At elevated temperatures, the resonance fields depicted in the right frames of Fig.~\ref{SpectraLT} yield a $g$ value ($g\approx1.98$) typical for Cr$^{3+}$ (electronic configuration $3d^3$, spin $S = 3/2$) in octahedral O$^{2-}$ coordination. On decreasing temperature the resonance line slightly shifts to lower resonance fields, strongly broadens, and disappears on passing $T_{\rm N}$, as illustrated in the left frames of Fig.~\ref{SpectraLT}. On approaching $T_{\rm N}$, where the linewidth reaches values of the same order of magnitude like the resonance field, it was necessary to take into account the counter resonance at $-H_{\rm res}$ for a reliable fitting as described Ref.~\onlinecite{Joshi2004}.
The double integrated intensity
\begin{equation}
I_{\rm ESR} = \int_{0}^{\infty}P(H)dH = \frac{\pi}{2} A \cdot \Delta H^2
\end{equation}
depicted in the left frames of Fig.~\ref{Spectra}, follows the static susceptibility at elevated temperature, but significantly deviates and decreases on approaching $T_{\rm N}$. This is different from the typical behavior in antiferromagnets, where the intensity follows the static susceptibility as long as the paramagnetic signal can be
resolved, like e.g. in the case of LiCuVO$_4$ in Ref.~\onlinecite{Krug2002}.

\begin{figure}
\centering
\includegraphics[width=75mm]{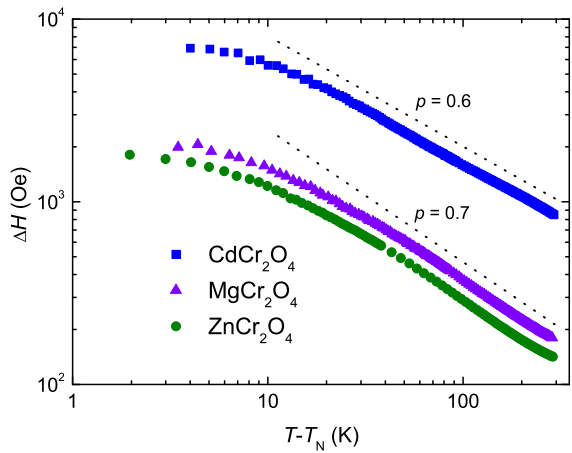}
\caption{(Color online) Temperature dependence of the ESR linewidth of MgCr$_{2}$O$_{4}$, ZnCr$_{2}$O$_{4}$, and CdCr$_{2}$O$_{4}$ as function of the reduced temperature $T-T_{\rm N}$ on a double logarithmic plot. Dotted lines indicate critical divergences $\Delta H \propto (T/T_{\rm N}-1)^{-p}$.}
\label{Linewidth2}
\end{figure}

Figure~\ref{Linewidth2} shows the ESR linewidths of all three compounds as a function of reduced temperature $(T-T_{\rm N})$ on a double logarithmic scale. Using the experimental values of $T_{\rm N}$ determined from the susceptibility, the data (solid symbols) monotonously increase on decreasing temperature but level off on approaching $T_{\rm N}$.
Only at elevated temperatures the data suggest a critical divergence $\Delta H \propto (T/T_{\rm N}-1)^{-p}$ with exponents $p < 1$ as indicated by the dotted lines. Such small critical exponents are far below the value ($p = 1.7$) expected for 3D Heisenberg antiferromagnets. Similar small values have been obtained in 2D triangular antiferromagnets $A$CrO$_{2}$.\cite{Hemmida2009,Hemmida2011} These compounds reveal a dominant Heisenberg character with a strong direct exchange interaction between Cr$^{3+}$ ions. Remarkably in three-dimensional (3D) pyrochlore chromium oxides, a similar dominant direct exchange interaction between Cr$^{3+}$ ions is also present.\cite{Yaresko2008}


Furthermore the deviation of the ESR intensities from static spin susceptibilities illustrated in Fig.~\ref{Spectra}, similarly, has been observed in the geometrically frustrated 2D chromium based ordered rock salts and delafossites. It has been attributed to the increasing influence of non-resonant relaxational modes with strongly increasing linewidth.\cite{Martinho2001} In 2D $A$CrO$_{2}$ these modes have been related to the $Z_{\rm 2}$ vortices.\cite{Hemmida2011} It seems that in $A$Cr$_{2}$O$_{4}$ such non-resonant relaxational modes can be associated with string modes, \textit{i.e.} the spin molecules mentioned above in the Introduction for the following reason: as has been indicated e.g. by Moessner and Ramirez \cite{Moessner2006} or Ross \textit{et al.},\cite{Ross2009} the geometrical frustration within the pyrochlore lattice strongly promotes 2D correlations on the (111) kagome planes inherent in this lattice structure. Especially the formation of the spin hexamers takes place exactly on these kagome planes. Therefore, we suggest that the spin-relaxation on the pyrochlore lattice is of dominantly 2D character indicating the tendency to undergo a topological phase transition analogous to the Berezinskii-Kosterlitz-Thouless (BKT) transition in the 2D XY magnet or the Kawamura-Miyashita (KM) scenario in the 2D triangular Heisenberg antiferromagnet.

Assuming this analogy, we adopted the theoretical framework of BKT vortices where the space $(\mathbf{r})$ and time $(t)$ dependent two-spin correlation function above the topological phase transition reads \cite{Mertens1989}
\begin{equation}
\langle S_x(0,0) S_x(\mathbf{r},t) \rangle \approx \frac{1}{2}S^2 \exp\{-\sqrt{\xi^2 \mathbf{r}^2 - \gamma^2 t^2}\}
\label{correlation}
\end{equation}
with spin $S$, vortex correlation length $\xi$, and $\gamma = \sqrt{\pi}\bar{u}/(2\xi)$, where $\bar{u}$ denotes the average vortex velocity. Fourier transformation yields the dynamic structure factor
\begin{equation}
S_{xx}(\mathbf{q},\omega) =  \frac{S^2}{2\pi^2}\frac{\gamma^3 \xi^2}{\{\omega^2+\gamma^2[1+(\xi q)^2]\}^2}
\label{DSF}
\end{equation}
probed by inelastic neutron scattering. In general the ESR linewidth is determined by four-spin correlations. Following Benner and Boucher,\cite{Benner1990} in independent-mode approximation, these four-spin correlations can be factorized into two-spin correlation functions. Taking into account that the microwave energy of 9~GHz is small compared to the energy scale of the vortices, the ESR linewidth probes the dynamic structure factor at zero momentum transfer $\mathbf{q}=0$ and nearly zero frequency $\omega \approx 0$ yielding
\begin{equation}
\Delta H \propto S_{xx}(\mathbf{q}=0,\omega \rightarrow 0) \propto \frac{\xi^2}{\gamma}
\label{ESR_DSF}
\end{equation}
Assuming the averaged vortex velocity $\bar{u}$ to be temperature independent, one finally arrives at $\Delta H \propto \xi^{3}$, i.e. the linewidth is determined by the vortex correlation length only.\cite{Becker1996}

Using this approximation, we tried to describe the temperature dependence of the resonance linewidth in the chromium-oxide spinels using the generalized vortex-correlation length $\xi$ derived by Bulgadaev \cite{Bulgadaev1999} as
\begin{align}
\xi &= \xi_{\rm 0}\exp[\frac{b}{(T/T_{\rm B}-1)^{\nu}}],\hspace{0.5cm} T>T_{\rm B}\nonumber
\\ \xi &= \infty,\hspace{0.5cm} T<T_{\rm
B} \label{Vortex Correlation length-1},
\end{align}
where $T_{\rm B}$ is the Bulgadaev temperature which substitutes the Kosterlitz-Thouless temperature in the BKT scenario. The exponent $\nu$ provides information on the internal symmetries of the spin system under consideration. For the pure BKT scenario of the XY magnet one obtains $\nu=0.5$. For the KM scenario of the triangular antiferromagnet, theoretical calculations derived $\nu \approx 0.4$ in agreement with experiment.\cite{Hemmida2011}

\begin{figure}
\centering
\includegraphics[width=70mm]{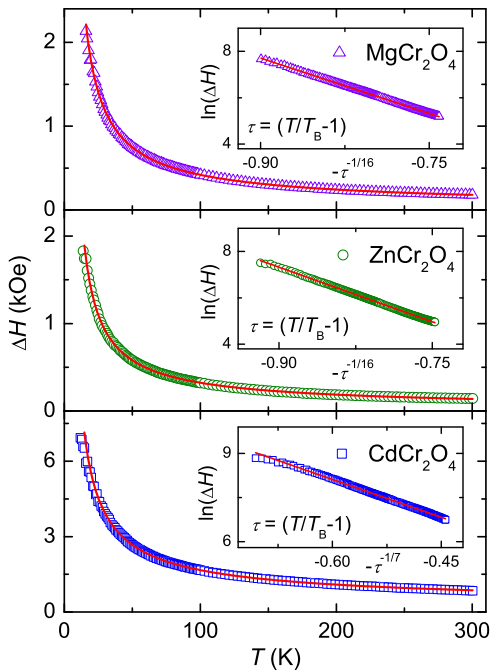}
\caption{(Color online) Temperature dependence of the linewidth $\Delta H$ of
MgCr$_{2}$O$_{4}$, ZnCr$_{2}$O$_{4}$, and CdCr$_{2}$O$_{4}$. The solid lines indicate fits by $\Delta H \propto \xi^3$ (Ref. \onlinecite{Becker1996}) using Eq.~\ref{Vortex Correlation length-1}. Insets:
Logarithmic plots of $\Delta H$ as ln$(\Delta H)$ \textit{vs} $-\tau^{-\nu}$, where $\nu$ values are $0.063\approx1/16$ for MgCr$_{2}$O$_{4}$ and ZnCr$_{2}$O$_{4}$, and $0.143\approx1/7$ in the case of
CdCr$_{2}$O$_{4}$.}
\label{Linewidth1}
\end{figure}

As shown in Fig.~\ref{Linewidth1}, Bulgadaev's scenario turns out to describe successfully the linewidth data $\Delta H \propto \xi^3$ in the full temperature range without any residual contributions. Only the exponent $\nu$ has to be chosen much smaller than those found in the 2D Heisenberg antiferromagnet so far.\cite{Hemmida2009,Hemmida2011} For the Mg and Zn compound $\nu\approx0.063$ yielded the optimum fit while for CdCr$_{2}$O$_{4}$ $\nu\approx0.143$ was preferable. The Bulgadaev temperature $T_{\rm B}$ is obtained as $2.47$, $2.82$, and $0.91\pm 0.05$~K, where $b=$ $5.29$, $5.27$ and $2.94\pm 0.05$ for $A=$ Mg, Zn, and Cd, respectively.


\section{Discussion}

First of all, it is important to note that the Bulgadaev temperature $T_{\rm B}$ exhibits values significantly below the N\'{e}el temperature $T_{\rm N}$ in all three compounds. This is very similar to the findings in most of the 2D triangular antiferromagnets $A$CrO$_2$, where the characteristic temperature $T_{\rm V}$ derived from the vortex scenario is located clearly below the magnetic ordering temperature.\cite{Hemmida2009,Hemmida2011} In those 2D frustrated antiferromagnets, this observation could be ascribed to the existence of a broad spin-fluctuation regime for $T_{\rm V} \leq T \leq T_{\rm N}$, where magnetic vortices coexist with antiferromagnetic order, as was confirmed by complementary experimental methods: for example, in NaCrO$_2$ ($T_{\rm N} = 41$~K) the characteristic temperature of $T_{\rm V} = 24$~K derived from the temperature dependence of the ESR linewidth was found to be in good agreement with the maximum of the broad peak of the $\mu$SR-relaxation rate.\cite{Olariu2006} Such a $\mu$SR-relaxation peak results from strong spin-fluctuations below $T_{\rm N}$. This was further supported by neutron scattering in NaCrO$_2$, detecting long-range order within the triangular planes,\cite{Soubeyroux1979} but only a weak incommensurate modulation along the $c$ axis.\cite{Hsieh2008} Taking into account the similarity of the observed relaxation scenario in 2D $A$CrO$_2$ with the 2D melting transition from a well-ordered solid into a disordered liquid via an intermediate liquid-crystal phase,\cite{Halperin1978, Young1979} the temperature regime $T_{\rm V} \leq T \leq T_{\rm N}$ can be regarded as magnetic analogue of this intermediate state.
Transferring these ideas to the chromium-oxide spinels, the temperature dependence of the linewidth suggests the existence of a wide spin-fluctuation regime $T_{\rm B} \leq T \leq T_{\rm N}$ for all three compounds where $T_{\rm B}$ is located more than $80\%$ below $T_{\rm N}$. Indeed, the existence of spin molecules above $T_{\rm N}$ in the correlated paramagnetic regime and the coexistence of the spin molecules with long-range antiferromagnetic order below $T_{\rm N}$, as detected by neutron scattering, indicate the importance of spin-fluctuations also below $T_{\rm N}$. However, if there is really another phase transition at $T_{\rm B}$, cannot be stated at present, because our ESR data are obtained in the paramagnetic regime and, therefore, only suggest the tendency for the behavior below $T_{\rm N}$.

Focusing on the nature of exponents $\nu\approx0.063$ and $\nu\approx0.143$, we realized that the obtained values are close to $\nu=1/16$ and $\nu=1/7$, respectively. According to Bulgadaev (\textit{cf.} Eq.~\ref{Coxeter} in Appendix), these exponents correspond to the so called Coxeter numbers $h_{G}=30$ and $12$, respectively. The number $h_{G}=30$ can indicate the Lie groups $A_{n=29}$, $D_{n=16}$ or $E_{8}$, while $h_{G}=12$ may indicate $A_{n=11}$, $D_{n=7}$ or $E_{6}$. In the following we provide hints based on recent experimental findings in literature that the exceptional groups $E_{8}$ and $E_{6}$ are probably relevant for the observed relaxation behavior.

In 3D the perfect realization of $E_8$ symmetry is the icosahedron. Therefore, icosahedral symmetry or -- its dual counterpart -- dodecahedral symmetry is expected to indicate the hidden $E_8$ symmetry. In this respect we work out indications for icosahedral or dodecahedral features appearing from different experimental results in $A$Cr$_2$O$_4$, especially focusing on analogies to icosahedral short-range order patterns in glasses.

(i) Dynamical molecular spin state: ultrasound velocity measurements in ZnCr$_{2}$O$_{4}$ and MgCr$_{2}$O$_{4}$ reveal elastic anomalies in the paramagnetic phase due to geometrical frustration.\cite{Watanabe2012} The trigonal shear modulus exhibits a characteristic minimum at $T\approx50$~K, while the temperature dependent tetragonal shear modulus shows huge Curie-type softening. The phonon frequency is said to soften, because its value is determined by restoring forces that soften on cooling.\cite{Dove2003} The soft modes are torsional (transverse) phonons which are slowing down the dynamical processes at a second-order critical point, i.e. at $T_{\rm N}$. Following Ref.~\onlinecite{Watanabe2012}, the behavior of the shear moduli strongly suggests the coexistence of a dynamical spin-Jahn-Teller effect, evident from the softening of the tetragonal shear modulus, and a dynamical molecular-spin state, related to the only weakly temperature dependent trigonal shear modulus, in the paramagnetic phase. This is compatible with the coexistence of magneto-structural order with cooperative tetragonal lattice distortion and a dynamical molecular-spin state with local trigonal lattice fluctuations in the antiferromagnetic phase. Thus, the magneto-structural ordering is explained by the spin-Jahn-Teller mechanism via magneto-elastic coupling, where cooperative distortions of the tetrahedra release the frustration in the nearest-neighbor antiferromagnetic interactions. Simultaneously, INS studies in MgCr$_2$O$_4$ revealed the appearance of gapped molecular-spin excitations in the antiferromagnetic phase, which corroborate the coexistence of magneto-structural order and zero-point motion-like fluctuations of the spin molecules.\cite{Tomiyasu2013}



(ii) Boson peaks: further analysis of INS results in ZnCr$_{2}$O$_{4}$ and MgCr$_{2}$O$_{4}$ concluded that broad spin-excitation peaks are usually signatures of short-range order.\cite{Lee2000,Tomiyasu2013} They are analogues to the short-range vibrational modes - so called boson peaks - observed in supercooled liquids and glasses.\cite{Grigera2003,Shintani2008,Frick1995} These boson peaks in glasses are equivalent to the transverse acoustic van Hove singularities in the vibrational density of states in crystals.\cite{Chumakov2011} In ZnCr$_{2}$O$_{4}$, van Hove singularities have been observed as a wave-vector degeneracy in the spin-wave spectra.\cite{Lee2000} The relation between the broad spin-excitation peaks and van Hove singularities can be clarified in the following way: geometrical frustration leads to constant energy surfaces or volumes for spin-wave dispersion relations in reciprocal spaces. Such Q-space "degeneracy" in turn yields pronounced van Hove singularities in wave-vector averaged spectra. A real-space interpretation has yet to be found for dispersionless excitations in the pyrochlore lattice. The broad peak indicates that they are highly localized in the ordered phase of ZnCr$_{2}$O$_{4}$.\cite{Lee2000} This is supported by theoretical calculations approving that, due to spin loops, van Hove singularities are seen in the local density of spin waves of ZnCr$_{2}$O$_{4}$.\cite{Tchernyshyov2002}


(iii) Dodecahedral spin clusters: interestingly, by applying INS methods in transition-metal-based quasi-crystals, Sato \textit{et al.} found that broad inelastic peaks can be interpreted as localized collective fluctuations of short-range ordered spins in dodecahedral spin clusters.\cite{Sato2006,Sato2008} Hence, they indicated a possible close relation between these peaks and boson peaks in glasses. By extending the Sethna-Sachdev-Nelson formula, \cite{Sethna1983,Sachdev1984} Kanazawa introduced a generalized view of the physical origin of the boson peak in the gauge-invariant formula.\cite{Kanazawa2002} He pointed out that the localized modes (massive gauge modes), which correspond to the boson peak, are required naturally through the Higgs mechanism. This means that localized modes, van Hove singularities, and spin clusters have a common nature.

Taking into account the analogy of short-range order fluctuations and cluster formation in frustrated spin systems, metallic glasses, and quasi-crystals, we suppose that the spin or molecular clusters in ZnCr$_{2}$O$_{4}$ and MgCr$_{2}$O$_{4}$ exhibit a hidden icosahedral symmetry, precisely the $E_{8}$-symmetry indicated by $\nu = 1/16$, which results in some kind of short-range order. This conjecture is based on the defect description of liquids and metallic glasses which are governed by an icosahedral order parameter.\cite{Nelson1983} Further experimental and theoretical efforts are necessary to check these ideas.

The result in CdCr$_{2}$O$_{4}$ raises additional questions. The $E_{6}$-symmetry derived from $\nu = 1/7$ would result in a tetrahedral configuration. However, neutron-scattering studies in CdCr$_{2}$O$_{4}$ did not reveal any kind of spin loops, probably due to the incommensurate spin order and its consequences.\cite{Chung2005,Matsuda2007} It seems that in CdCr$_{2}$O$_{4}$ bond frustration due to super exchange interactions gains more influence as compared to the purely geometrically frustrated ZnCr$_{2}$O$_{4}$ and MgCr$_{2}$O$_{4}$ in which the direct nearest-neighbor (Cr$^{3+}$-Cr$^{3+}$) exchange interaction dominates. For example, tetramers and di-tetramers were detected in GeCo$_{2}$O$_{4}$ in which the ferromagnetic Co$^{2+}$-O$^{2-}$-Co$^{2+}$ super exchange plays an important role.\cite{Tomiyasu2011c,Watanabe2011} Probably this tendency influences the magnetic behavior in CdCr$_{2}$O$_{4}$.

\section{Conclusion}

To summarize, we found that the spin-spin relaxation mechanism in $A$Cr$_{2}$O$_{4}$ follows a Beresinskii-Kosterlitz-Thouless like scenario similar to the 2D triangular layered antiferromagnets $A$CrO$_{2}$ but with significantly smaller exponent $\nu$. In case of strongly geometrically frustrated MgCr$_{2}$O$_{4}$ and ZnCr$_{2}$O$_{4}$ the value $\nu = 1/16$ was obtained from the temperature dependence of the ESR linewidth. Based on Bulgadaev's model of a generalized BKT transition with internal symmetry and considering the analogy to metallic glasses and quasi-crystals, this exponent can be ascribed to the exceptional Lie group $E_{8}$ indicating a hidden icosahedral symmetry of the spin molecules formed on the frustrated lattice.

\acknowledgments

We thank Dana Vieweg for the susceptibility measurements. This work was supported by the Deutsche Forschungsgemeinschaft (DFG) within the Transregional Collaborative Research Center TRR 80 (Augsburg-Munich-Stuttgart).

\appendix

\section{}
There are several ways of introducing symmetry. In mathematics, symmetries are usually associated with operations (translations, rotations, and reflections) that leave a geometrical object invariant. The collection of such operations forms a mathematical group. The mathematical description of continuous symmetries (as opposed to discrete symmetries, such as those that leave a crystal lattice invariant) is codified in the notation of a Lie group. There are nine Lie groups which are classified into two families, classical and exceptional. The first one includes four different groups known as $A_{n}=SU(n+1)$, $B_{n}=SO(2n+1)$, $C_{n}=SP(2n)$, $D_{n}=SO(2n)$, of respective $n\geq1$, $n\geq2$, $n\geq3$, $n\geq4$.\cite{Itzykson1989,Penrose2007} The subscript $n$ is called the rank of the group and measures how large the group is. There is no restriction of its maximum value. The notations $SU$, $SO$, and $SP$ refer to special unitary, special orthogonal, and symplectic, respectively. The second family includes five groups which are known as $G_{\rm 2}$, $F_{\rm 4}$, $E_{\rm 6}$, $E_{\rm 7}$, and $E_{\rm 8}$. The words classical and exceptional are used in a sense that one knows, how to describe the classical groups for every $n$. In contrast, in the exceptional groups, although the index value will not exceed the value $8$, these groups have very high symmetries, which cannot be obeyed by any familiar geometrical object.

On the other hand, the objects of highest symmetry in three-dimensional space are regular polyhedra which are well-known as Platonic solids: the tetrahedron, cube (hexahedron), octahedron,
dodecahedron, and icosahedron. The cube and octahedron are "dual" to each other, the same holds for the icosahedron and the dodecahedron. The word "dual" means that one can get one Platonic solid from the
other when the center of each face becomes a vertex of the dual. Dual solids have the same symmetry group, so there are three symmetry groups here: the tetrahedron, cube, and icosahedron. These three symmetry groups are the perfect realization of exceptional Lie groups $E_{\rm 6}$, $E_{\rm 7}$, and $E_{\rm 8}$ in 3D. Mathematicians project hyper-dimensional geometrical objects into the 2D plane. Coxeter derived a theory of projection of hyper-space objects into the plane.\cite{Coxeter1973} In this way, mega-symmetry operations as rotations, reflections \textit{etc.} are tightly packed.

Including all possible internal symmetries of 2D systems into the description of BKT phase transitions, Bulgadaev \cite{Bulgadaev1999} derived the vortex-correlation length $\xi$ given in Eq.~\ref{Vortex Correlation length-1}. The exponent $\nu$ is connected to one of the polytope characteristics, the so-called Coxeter number $h_{G}$ in the following way:
\begin{equation}
\nu = \frac{2}{2+h_{G}}\label{Coxeter}
\end{equation}
where the subscript $G$ refers to the class of the group $G$ that describes the packed local symmetry in the lattice. The Coxeter number describes the reflection operations in the symmetry group.


\end{document}